# Electrospun light-emitting nanofibers as building blocks for photonics and electronics

Andrea Camposeo, Luana Persano and Dario Pisignano

*Nanomaterials made of active fibers have the potential to become new functional components of light-emitting sources in the visible and near-IR range, lasers, and electronic devices*

Organic semiconductors and light-emitting materials have stimulated significant research in the last two decades. Starting with the discovery of electroluminescence in conjugated polymer materials,[1] this research has led to the demonstration of a large variety of plastic and flexible devices for photonics and electronics. Many of these, particularly displays based on organic light-emitting devices, have already found their way into the market. Today, active polymer nanomaterials continue to be a very active field of research driven by the interest in pushing the miniaturization limits of light sources, in building sub-wavelength optical components, and in seeking exotic photonic effects in nanostructured systems and lasers.[2]

In this framework, polymer nanofibers realized by electrospinning deserve particular attention.[3,4] Due to their ease of fabrication, low cost, and chemical and compositional flexibility, these nanostructures have the potential to become functional building blocks of organic photonics and electronics. In addition, the relatively high throughput of their fabrication method means that many of these nanomaterials are easily available at laboratory scale and can be industrially exploited.[5]

Researchers initially developed electrospinning to realize nanofibers from optically inert polymers, which usually exhibit good processing and viscoelastic behavior.[4,6] This technique is based on the uniaxial elongation of a viscous polymer solution following the application of a voltage bias between a metallic needle, which is at the end of the syringe that contains the liquid, and a collector. Depending on the polymer solution and processing parameters used, the average diameter of the resulting fibers can range from ∼10 nm to ∼10 μm, thus covering the dimensions of interest for electronics and photonics in the visible and near-IR.

The fabrication process of these nanofibers is operationally straightforward, but it may be complicated by the large number of experimental parameters involved. These include parameters related to the polymeric solution used, the experimental setup, and the environment where the process is performed (such as temperature and humidity). The solvents employed, specifically their viscoelastic properties, play a major role in determining the outcome of the fabrication step. In particular, light-emitting materials such as many conjugated polymers are often difficult to spin because they can exhibit poor viscoelasticity. The strategies allowing these materials to be electrospun include a proper choice of the solvents used and a very careful optimization of other parameters such as the solution concentration and chemical composition.

Recently, our group realized light-emitting nanofibers based on conjugated polymers. We also demonstrated the feasibility of patterning the nanofibers' surfaces with features at the 100nm scale with a room-temperature method that did not degrade their emission.[7,8] Using these and other organic active materials (see Figure 1), we realized devices such as lasers,[9] phototransistors,[10] field effect transistors,[11] and polarized light sources embedded in lab-on-a-chip devices.[12] Further, we found a range of interesting effects such as emission over a wide range of wavelength and Förster energy transfer in the fibers,[13, 14] which demonstrates a high degree of achievable spectral tunability. Another interesting characteristic of our polymer nanofibers is an enhanced efficiency compared to thin films for some conjugated co-polymer species,[15] which is indicative of a peculiar supramolecular organization compared to conventional systems. Furthermore, when we produced nanofibers using organic materials that exhibit optical gain, we typically observed lasing under optical pumping for fluences of the order of $10^2$ μJcm$^{-2}$. This means that these nanostructures can constitute compact and very cheap lasing media.

The flexibility of electrospinning is such that it enables significant variation in terms of processing and achievable nanostructures. For instance, it is possible to obtain more uniform fibers by adding organic salts to the electrospun solution. Importantly, we have found that the presence of these salts does not significantly degrade the optical properties of conjugated polymers.[16]

Active nanofiber-based systems that can be electrospun do not include only emissive and conductive materials. With some polymers and a proper choice of solvents, we obtained free-standing arrays in which nanofibers slightly overlap, forming mutual joints that increase the mechanical robustness of the overall produced material: see Figure 2. In this way, we and our collaborators realized piezoelectric textiles usable for building sensors with unprecedented pressure sensitivity performance.[17]

In summary, polymer nanofibers are a novel class of organic nanostructures that show excellent chemical and functional flexibility. As a result, they can be used as active elements in light-emitting sources, lasers, energy harvesters, and sensors. We anticipate the development of these devices to implement hybrid functionalities interplaying with optical and lasing properties. More generally, the availability of a portfolio of technical methods to realize bright light-emitting nanofibers makes the process particularly appealing for future studies. These include a more in-depth investigation of the fundamental features of electrospinning, which are not yet completely understood for active polymer solutions. The aim of this future work is to link the resulting optical and lasing properties to the complex multi-physics leading to the fiber formation.

*The research leading to the most recent of these results[16,17] has received funding from the European Research Council (ERC) under the European Union's Seventh Framework Programme through ERC Grant Agreement 306357 (ERC Starting Grant NANO-JETS).*

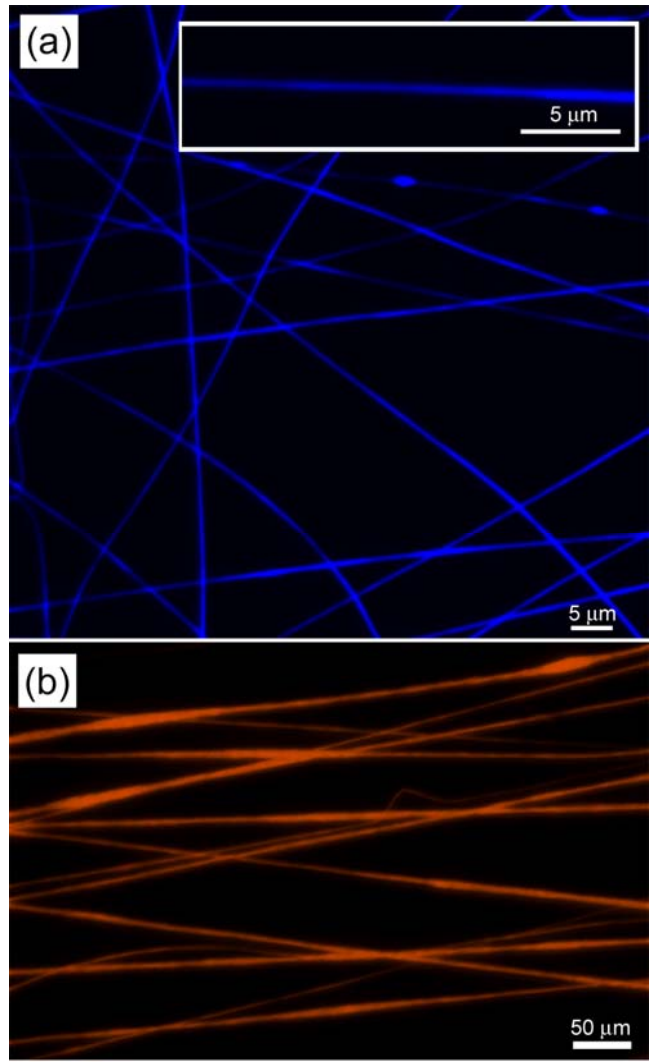

**Figure 1.** *Confocal fluorescence micrographs of light-emitting, electrospun fibers based on (a) poly-[(9,9-dioctylfluorenyl-2,7-diyl)-co-(N,N'-diphenyl)-N,N'-di(p-butyl-oxy-phenyl)-1,4-diaminobenzene) and (b) poly[2-methoxy-5-(2-ethylhexyl-oxy)-1,4-phenylene-vinylene]. Inset: Magnification of a single active fiber.*

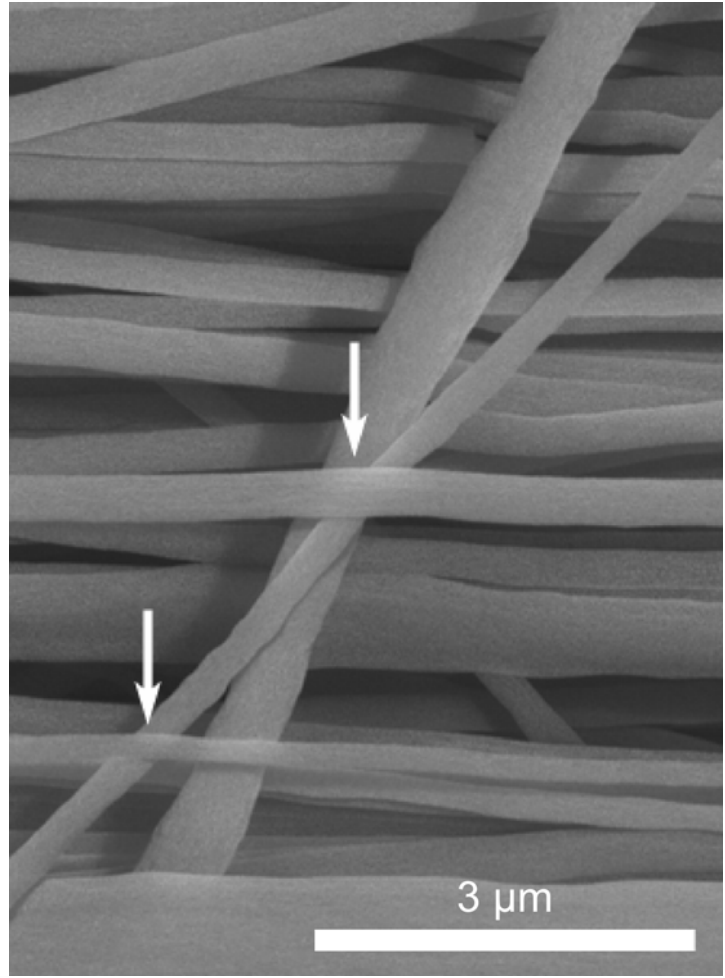

**Figure 2.** *Scanning electron microscopy image of electrospun active fibers showing mutual joints (indicated by arrows) obtained during electrospinning that make the overall material mechanically robust. The arrows indicate examples of joints in the micrograph.*

**Author Information**

**Andrea Camposeo, Luana Persano**
National Nanotechnology Laboratory (NNL)
Nanosciences Institute, CNR
Lecce, Italy

Andrea Camposeo is a researcher working on the development of light-emitting polymer nanofibers for photonics and organic-based laser devices. His interests also include characterization of the emission and gain properties of polymers, composites, and organic crystals, and two-photon lithography.

Luana Persano is a researcher working on nanofabrication on soft matter, piezoelectric nanofibers, soft lithography, nanocomposite materials, and organic-based laser devices. She has received the CNR Start Cup Award and the Bellisario Prize for a Young Talent in Industrial Engineering for her work on electrospinning technology transfer.

**Dario Pisignano**
Department of Mathematics and Physics 'Ennio De Giorgi'
University of Salento
and
NNL
Nanoscience Institute, CNR
Lecce, Italy

Dario Pisignano is an associate professor of experimental physics and nanobiotechnologies, the coordinator of the Soft Matter Nanotechnology Group at NNL, and principal investigator of the NANO-JETS ERC Starting Grant Project. His research interests include polymer nanofibers, photonic devices based on organic materials, microfluidics, and nanobiotechnology.